\documentclass{article}

\usepackage{arxiv}

\usepackage[utf8]{inputenc} 
\usepackage[T1]{fontenc}    
\usepackage{hyperref}       
\usepackage{url}            
\usepackage{booktabs}       
\usepackage{amsfonts}       
\usepackage{nicefrac}       
\usepackage{microtype}      
\usepackage{lipsum}		
\usepackage{graphicx}
\usepackage[numbers]{natbib}
\usepackage{doi}
\usepackage{color,soul}
\usepackage{placeins}
\newcommand{\statistic}[1]{\textcolor{black}{#1}}

\title{
American postdoctoral salaries do not account for growing disparities in cost of living
}

\author{Tim Sainburg \\
	Harvard Medical School \\
	UC San Diego \\
	\texttt{tim\_sainburg@hms.harvard.edu} \\
}

\renewcommand{\headeright}{}
\renewcommand{\undertitle}{}

\begin{document}
\maketitle

\begin{abstract}
The National Institute of Health (NIH) sets postdoctoral (postdoc) trainee stipend levels that many American institutions and investigators use as a basis for postdoc salaries. 
Although salary standards are held constant across universities, the cost of living in those universities' cities and towns vary widely. 
Across non-postdoc jobs, more expensive cities pay workers higher wages that scale with an increased cost of living. 
This work investigates the extent to which postdoc wages account for cost-of-living differences. 
More than 27,000 postdoc salaries across all US universities are analyzed alongside measures of regional differences in cost of living. 
We find that postdoc salaries do not account for cost-of-living differences, in contrast with the broader labor market in the same cities and towns. 
Despite a modest increase in income in high cost of living areas, real (cost of living adjusted) postdoc salaries differ by \statistic{29\%} (\statistic{\$15k} 2021 USD) between the least and most expensive areas. 
Cities that produce greater numbers of tenure-track faculty relative to students such as Boston, New York, and San Francisco are among the most impacted by this pay disparity. 
The postdoc pay gap is growing and is well-positioned to incur a greater financial burden on economically disadvantaged groups and contribute to faculty hiring disparities in women and racial minorities.
\end{abstract}


\section{Introduction}

American funding agencies typically do not account for geographic differences in cost of living when setting salary standards. For example, the National Institute of Health (NIH), the largest biomedical research funding agency in the US, sets a constant rate of pay for all postdocs awarded NRSA fellowships and restricts on how income can be supplemented for fellows \cite{nih_policy}. 
Although the NIH payscale does not restrict non-fellowship salaries, many US institutions use NIH NRSA standards as a basis for their own postdoctoral salary levels.

The same salary can vary greatly in America across economic regions in terms of the goods and services (e.g. rent, food, transportation) that it can be used to purchase. The US Bureau of Economic Activity (BEA) estimates this difference across geographic regions as the Regional Price Parity (RPP) which is used as the basis of determining an individual's "real" income. Real income estimates the value of an individual's income, adjusted by setting the value of dollars constant across geographic regions using the RPP \cite{aten2019regional}. 
For example, the 2021 first-year postdoc NRSA stipend, set by the NIH at \statistic{\$53,760}, provides a real salary of \statistic{\$56,709} in Durham, North Carolina (RPP=\statistic{94.8}), and only \statistic{\$39,970} in San Francisco (RPP=\statistic{134.5}). 

Although larger cities are typically more expensive places to live, wages typically scale with the cost of living \cite{glaeser2001cities, winters2009wages}. 
This phenomenon is explained by the economic notion of spatial equilibrium, which predicts compensatory equality between income, prices, and amenities \cite{winters2009wages, glaeser2008cities}. 
Because funding agencies set uniform postdoctoral salary standards across institutions without respect to the cost of living, we hypothesize that spatial equilibrium will not hold for postdoc salaries in the US, resulting in a salary disparity across American cities. 

Postdoc positions are generally considered to be a transitory period between training and finding a permanent academic position \cite{nerad1999postdoctoral}. 
The treatment of the postdoctoral career stage as a training period rather than long-term employment can lead to a disregard for postdoctoral salary as an important factor in choosing a position \cite{russo2008us, athanasiadou2018assessing}, despite the long-lasting financial impacts of postdoctoral training \cite{kahn2017impact}. 
The majority of tenure-track faculty at universities are trained by a small minority of universities that are often located in America's most expensive cities \cite{clauset2015systematic}. 
If higher-prestige universities pay lower real salaries, a short-term loss in income could still be a valuable investment into future career outcomes. 
However, a lower salary has the potential to disproportionately impact researchers who cannot absorb the short-term impacts of a lower postdoc income, such as parents and individuals from disadvantaged socioeconomic roots.

This work explores how income differs between postdocs when factoring in the cost of living across geographical regions. 
First, we find that, despite a modest increase in postdoc salaries in high cost of living areas, the value of a postdoc's income in the highest cost of living areas is \statistic{29\%} (\statistic{15k} 2021 dollars) lower than for postdocs living in the lowest cost of living areas. 
This effect is not observed in the general job market, where income scales with the cost of living. 
Second we find that the postdoc salary gap is particularly prominent in the cities and universities that produce the most tenure-track faculty. 
Finally, we find that these pay disparities have likely increased with changes in the cost of living across cities since the 1980s.
Taken in context, these data provide indications for a source of inequity in the academic pipeline.

\section{Results}
\label{sec:Results}

To compare salaries across institutions we use a dataset of \statistic{27,572} postdoctoral salaries acquired from H1B postdoctoral hires between 2015 and 2020 reported by the US Department of Labor. We adjust salaries to their 2021 equivalent for inflation across years to 2021 dollars using the Consumer Price Index (CPI), a measure of inflation. Although departmental information was not available for every salary datapoint, the most commonly observed positions were in biomedical disciplines. 

\subsection{Postdoc salaries do not fully account for cost of living}

\begin{figure}[!htb]
  \centering
  \includegraphics[width=1 \textwidth]{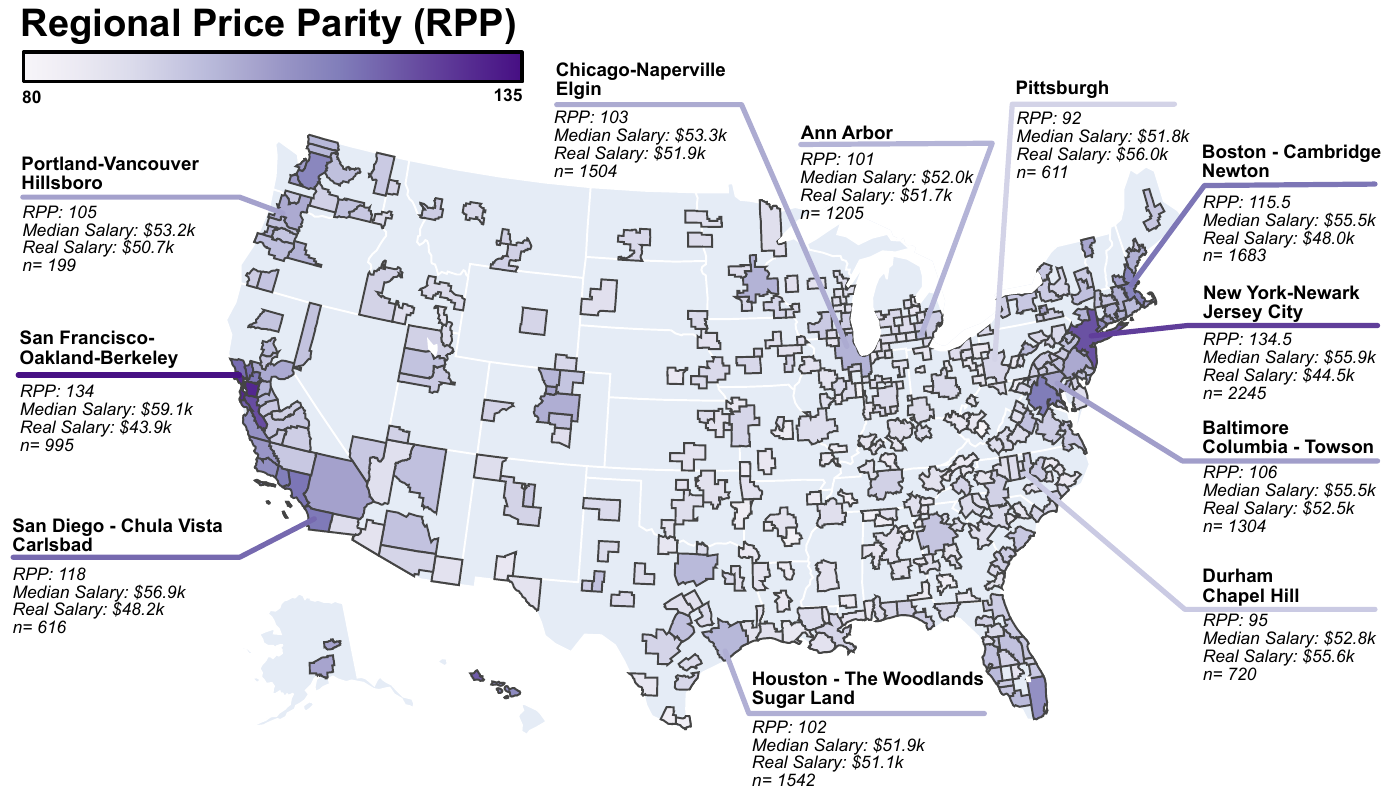}
  \caption{Regional Price Parity across CBSAs. Salaries listed refer to the median salaries and real salaries for postdocs in those regions. 
  }
  \label{fig:map} 
\end{figure}

We adjust postdoc salaries to their "real salary" based upon the Core-Based Statistical Area (CBSA; socioeconomically tied geographic regions) that the university resides within. Each year, the United States Bureau of Economic Analysis (BEA) releases the Region Price Parity (RPP) across US geographical regions, which measures the difference in price levels weighting rents, food, transportation, housing, recreation, education, medical, apparel, and other expenditures (Fig \ref{fig:map}). Using the RPP, the BEA estimates the Real Personal Income across the different geographical regions, which is a price-adjusted regional income. We apply the same transformation to postdoc salaries for each university and salary to estimate its real postdoc salary. Figure \ref{fig:map} shows the median postdoc salaries and median real postdoc salaries across several statistical areas with large postdoc populations. 

\begin{figure}[!htb]
  \centering
  \includegraphics[width=0.6 \textwidth]{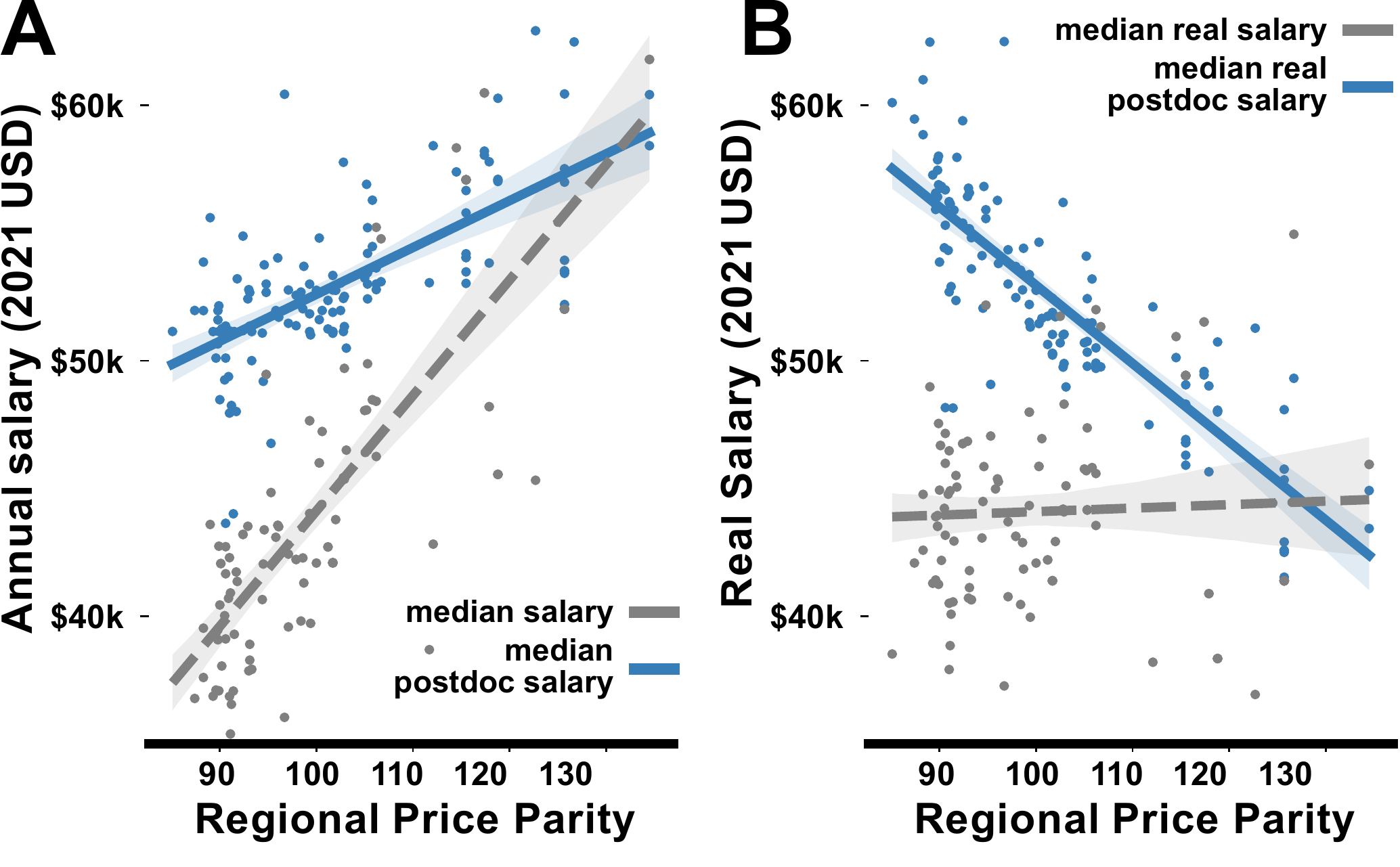}
  \caption{Salary adjustments for postdocs do not account for the cost of living. (A) The median postdoc salary for each university. Each blue point corresponds to the median salary for a single university. Each grey point corresponds to the median salary for the same CBSA that the university resides within. 
 (C) The same salary data as in (A) scaled with RPP to the real salary. 
  }
  \label{fig:col}
\end{figure}

As a comparison to postdoc salaries and real salaries we look at the median salaries across all jobs in the same CBSA reported by the US Bureau of Labor Statistics (BLS). Consistent with prior observations \cite{glaeser2001cities}, we find that salary across the job market scales with the cost of living, making real salary nearly constant \statistic{($r$(124)=0.04, $p$=0.63)} across different RPPs ((\ref{fig:col}A,B gray line). 
In contrast, postdoc salaries do not scale with the cost of living. While we see a modest increase in postdoc salaries in the most expensive cities (Fig \ref{fig:col}A blue line; e.g. \statistic{11\%} or \statistic{\$6,010} in San Francisco), real postdoc salaries are negatively correlated with RPP (Fig \ref{fig:col}B blue line; \statistic{$r$(124)=-0.85, $p$=5.17e-35}).

\subsection{Top tenure-track faculty producing universities pay lower real salaries}

\begin{figure}[!htb]
  \centering
  \includegraphics[width=1.0 \textwidth]{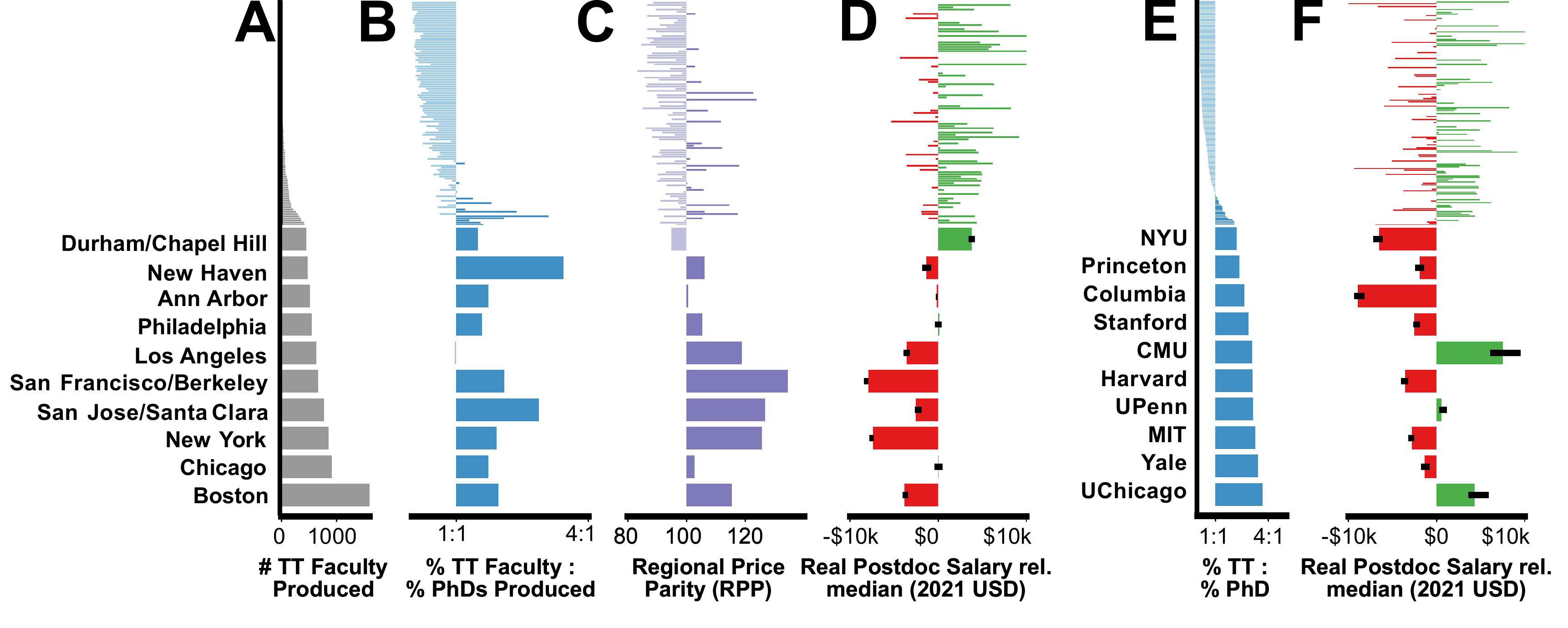}
  \caption{Postdoc salaries in the cities and universities producing the most postdocs. 
  (A) Cities sorted by number of TT faculty with a PhD from that city from \cite{clauset2015systematic}
  (B) Overrepresentation ratio of TT faculty produced to PhD students produced from (A). 
  (C) RPP for each city. 
  (D) Difference of median real postdoc salary for city relative to the national median (\statistic{\$51,849}) with 95\% CI.
  (E) Universities sorted by the ratio of TT Faculty to PhD students produced, as in (B). 
  (F) Difference of median real postdoc salaries for the universities in (E) relative to the national median (\statistic{\$51,849}). 
  }
  \label{fig:faculty}
\end{figure}

Despite lower pay, there is a benefit to be gained by doing a postdoc in America's most expensive cities. 
Where you train has an outsized impact on your ability to find a tenure-track (TT) faculty position \cite{clauset2015systematic}. 
The prestige of PhD and postdoc training institutions are among the strongest indicators of future positions \cite{mc1982postdoctoral, clauset2015systematic}. Faculty placements are also increasingly dependent on fellowships and career development awards which are unevenly distributed across institutions \cite{pickett2019increasing}.
Clauset et al.  \cite{clauset2015systematic}, created a dataset of TT faculty in Business, Computer Science, and History, documenting the universities where they received their PhD and found that just 25\% of universities produce 71-86\% of all tenure-track (TT) faculty hires. Here, we use Clauset et al.'s dataset (faculty at US universities n=\statistic{13,866}) to observe the relationship between postdoc wages, cost of living, and academic career prospects. Although this dataset reports data on PhD training institutions, postdoc training institutions play a similar or greater role in TT career prospects in biological sciences \cite{mc1982postdoctoral}.

As can be seen in Fig \ref{fig:faculty}A-D, we find that the American cities that produce the greatest number of TT faculty produce a greater share of TT faculty than trainees, are generally more expensive places to live, and pay postdocs lower real salaries. For example, the Boston metropolitan area produces the greatest share of TT faculty at \statistic{12}\% and only \statistic{6}\% of PhD trainees. Boston has an RPP that is \statistic{115.5}\% of the American average and the real salary of the median postdoc in the Boston area is only \statistic{93}\% of the American median. 

The same trend holds true at the level of the University. Of the universities that produce the greatest ratio of TT faculty to trainees (Fig \ref{fig:faculty}E), those in higher cost of living cities pay a lower real salary than average (Fig \ref{fig:faculty}F).

There are universities and cities that stand out. Chicago, for example, has an RPP only slightly above average (Fig \ref{fig:faculty}A-D), and the University of Chicago, which produces the greatest ratio of faculty to PhDs, pays a greater real salary than the median (Fig \ref{fig:faculty}E-F). The same is true of schools like Carnegie Mellon in Pittsburgh, and Duke in Durham.

\FloatBarrier

\subsection{Cost of living disparity is increasing across universities}

\begin{figure}[!htb]
  \centering
  \includegraphics[width=0.7 \textwidth]{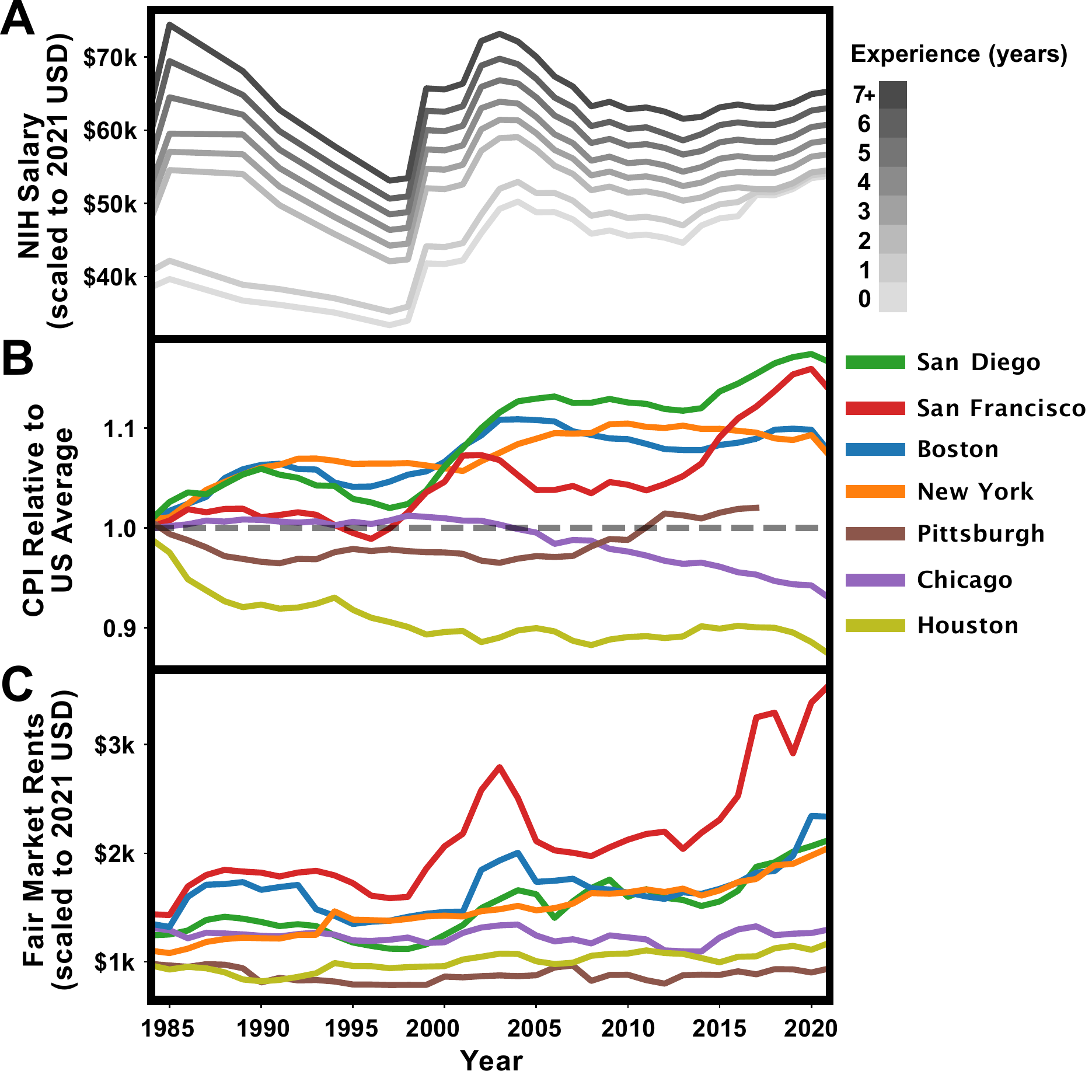}
  \caption{
 Changes in NIH postdoc salary standards relative to CPI across cities and Fair Market Rents between 1984 and 2021. NIH salaries and Fair Market rents are inflated to 2021 and show the price of a 2-bedroom apartment. 
  }
  \label{fig:time}
\end{figure}

Cost-of-living disparities between cities are not new. However, the-cost-of living gaps between many cities with large academic populations are increasing. In major university cities like Boston, San Francisco, and San Diego, the emergence of tech and biotech industries have gentrified and increased housing prices throughout the 1990s, 2000s, and 2010s \cite{sable2007impact}. 
To measure changes in disparities in the cost of living across cities in comparison to postdoc salaries we look at the Consumer Price Index 
and Fair Market Rents. 

NIH salaries since the 1980s have fluctuated year-by-year but remain in a similar range to what they are today (Fig \ref{fig:time}A). Postdoc salaries bottomed out in the mid-1990s and then peaked in 2003, where a postdoc with 7+ years of experience made a salary of \$73,145 (2021 USD), in comparison with the salary of \$65,282 in 2021. Over the past decade, postdoc wages have been steadily rising, with the greatest salary gains for early trainees. 

The Consumer Price Index (CPI) is a weighted average of the prices of consumer goods like food, medical care, housing, and transportation and is used as a measure of inflation. CPI measures changes in prices, but not absolute differences in prices. The BLS has released CPI values relative to the prices of goods in 1982-1984 since 1984 across US economic regions (Figure \ref{fig:time}B). The cost of goods in regions like Boston, San Diego, San Francisco, and New York City have raised relative to the national average over the past 40 years while cities like Houston and Chicago have seen decreases in prices relative to the national average (Fig \ref{fig:time}B). 

The Fair Market Rents is a statistic published by the U.S. Department of Housing and Urban Development, measured as the 40th percentile rents in metropolitan areas. Where CPI provides information about changes in prices, Fair Market Rents provide information about differences in housing prices across economic regions, a major factor in the cost of living and spatial equilibrium \cite{winters2009wages}. Differences in rent prices in these major metropolitan areas have increased since the early 1980s (Fig \ref{fig:time}C). For example, the monthly fair market rent of a 2-bedroom apartment in San Francisco was \statistic{\$1,430} 2021 dollars in 1984 compared to \statistic{\$3,553} in 2021.  

Taken together, postdoc salaries, as indicated by NIH standards, have fluctuated in the same range over the past four decades. Meanwhile, cost-of-living disparities across many academic cities have grown. 

\FloatBarrier
\section{Discussion}
\label{sec:Discussion}

American postdoc salaries are held relatively stable across the United States despite differences in cost of living across cities. These differences result in a lower real income for postdocs living in more expensive areas. In part this is by design: the NIH sets standards on postdoc stipends that are held constant across the US. This pay disparity is increasing. 
These differences in pay are reflected most strongly in the cities that produce the most TT faculty like Boston, San Francisco, and New York. 
These same cities are becoming increasingly expensive relative to the rest of the United States, resulting in a decrease in the real value of a postdoc's income. 

How do these pay differences impact the scientific landscape?  
Although substantial efforts have recently been put into place to reduce inequalities in the opportunities available to American scientists \cite{nih_almanac}, parity has not been reached, particularly across race, gender, and socioeconomic background \cite{nsf_state, morgan2021socioeconomic}. 
For example, women account for over half of US doctoral recipients in biology but comprise a much smaller proportion of the distribution of TT professors \cite{nsfsurvey, pell1996fixing, llorens2021gender, morgan2021unequal}. 
One factor impacting this disparity is an imbalance in the family-related responsibilities between women and men \cite{ceci2011understanding} which can be exacerbated by financial hardship such as an inability to pay for increasingly costly childcare \cite{lee2017parents}. 
The disparity between males and females in postdoc positions is greatest in laboratories that produce the largest number of tenure track professors \cite{sheltzer2014elite}, which are largely located in cities like Boston, New York, and San Francisco. 
It is unlikely that a reduction in income is the sole cause of the "leaky pipeline". 
However, decreased pay during postdoctoral training and the downstream effects on parenthood and work-life balance may unequally influence individuals who choose to pursue academic careers.
Similar issues arise for those who enter the academic path with college debt or without an available safety net allowing them to absorb short-term financial losses in order to pursue long-term career benefits. 
For example, student loan debt is greater in black students than white students in America \cite{jackson2013price}. These disparities do not fall in line with the NIH mission \cite{nih_almanac}.

Would anything be gained by the scientific community by paying commensurate salaries to postdocs in more expensive cities?
This question parallels one that pervades urban economics: why do businesses choose to locate in dense cities, where they have to pay higher wages? 
It is believed that businesses choose to pay more for workers in dense areas because proximity adds some degree of value through productivity \cite{glaeser2001cities}. 
It remains an open question whether or to what extent proximity between universities in areas like Boston, New York, or San Francisco aid in research productivity. Here, we do observe that many high cost of living cities produce larger numbers of TT faculty relative to trainees. 

Spatial equilibrium predicts a compensatory balance between wages, prices, and amenities. The analyses here compared wages and prices (e.g. rent, cost of living) but not amenities. Regions such as New England, the Pacific coast, and the Mid Atlantic have more highly valued amenities than the rest of the US, meaning workers are willing to accept a lower real wage to work there \cite{winters2009wages}. 
Still, our comparisons with non-postdoc salaries in the same cities show that postdoc salaries do not scale to the same extent. 
For example, in 2021 the salary of the median elementary school teacher in San Francisco was \statistic{70\%} greater than Durham/Chapel Hill, NC (\statistic{\statistic{\$82,870} vs \$48,620}), 
while the median salary of a postdoc in San Francisco was only \statistic{12\%} greater than in in Durham/Chapel Hill (\statistic{\$59,110} vs. \statistic{\$52,761}). 

Post-graduation, PhD students are faced with the decision to continue in the traditional academic trajectory or search for employment in other sectors like industry or government. 
The pay disparity for postdocs driven by the cost of living described in this work pales in comparison to pay disparities between postdocs and PhDs working in industry. 
The yearly NSF Survey of Earned Doctorates reported in 2020 that the median life sciences PhD in their first year of industry careers expects to make just over \$100k USD, while the median postdoc expects to make half of that, only \$50k \cite{nsfsurvey}. 
Within postdoc salaries, cost-of-living differences account for real salaries up to \$10k USD greater or less than the median real salary (\statistic{\$51,849}), a difference of close to \statistic{20\%}.

Taken together, the data presented here provide new insights on the pay disparities between postdocs in different regions of America and the direction those pay disparities have been moving over the past four decades. Salary differences may also be present in income at other stages in the academic career such as graduate student stipends and early career tenure-track researchers, where funding agencies also place caps on salaries. The consequences of these disparities are unknown and warrant further exploration.

\bibliographystyle{unsrtnat}
\bibliography{references}  
\newpage

\section{Methods}

\subsection{Datasets}
\label{sec:datasets}

\paragraph{H1B Postdoctoral salaries}
Postdoctoral salaries were acquired through available H1B applications through the US Department of Labor (DOL). We used available H1B data from 2015-2020 and subsetted postdoctoral jobs at universities. Yearly salary is given in the column \texttt{WAGE\_RATE\_OF\_PAY} or \texttt{WAGE\_RATE\_OF\_PAY\_FROM} depending on the year/dataset.

\paragraph{University locations}
University locations were taken from the US Department of Education's National Center for Education Statistics, which maintains a dataset of all US universities and their locations. The dataset can be found at \url{https://nces.ed.gov}.

\paragraph{Core-Based Statistical Areas (CBSA)}
Core-Based Statistical Areas were gathered from the US Census Bureau datasets for 2020. CBSAs comprise metropolitan and micropolitan statistical areas, defined by their populations (50k or more individuals for metropolitan, 10k or less for micropolitan), and adjacent territory that has a high degree of social and economic integration with the core, as measured by commuting ties. 
This dataset can be found at \url{https://www.census.gov/}. 

\paragraph{Regional Price Parity (RPP)}
Regional price parities (RPPs) are released yearly by the US Bureau of Economic Activity. RPPs measure differences in price levels across economic regions for a given year and are expressed as a percentage of the overall national price level. RPP data are available online through the BEA website
(\url{https://www.bea.gov/}).

From 2008 until 2021 RPP was computed on the basis of the Consumer Price Index (CPI). 
In December 2021 the BEA released an experimental new calculation for RPP, based upon Personal Consumption Expenditures (PCE). 
There are several differences in the new metric, including changes in the weighting scheme between e.g. rents, food, medical costs. 
The primary difference between the CPI and PCE metrics is that CPI price data and weights are designed to measure out-of-pocket expenditures for individuals, whereas PCE expenditures include third parties, such payments for an individual made by the government and insurance companies \cite{bea_cpi_pce}. 
We chose to use the CPI weighted RPP values rather than the PCE-based RPP because our interest is in the cost of living, which the new measure does not reflect. 
To compute real salaries we used RPPs from the most recent RPI-based estimate \statistic{2019 (released in December 2020)}. RPPs for universities in non-metropolitan areas were set at 86.8, the national nonmetropolitan RPP value for 2019. 

\paragraph{BLS salary data}
Salary data across occupations were taken from the US Bureau of Labor Statistics Occupational Employment and Labor Statistics (BLS OES \url{https://www.bls.gov/oes/}). We used the median salaries matched to the statistical area corresponding to each university. 

\paragraph{NIH NRSA salaries}
NIH Kirschstein-NRSA salaries are announced yearly. Since 1975 NRSA salaries have been set for predoctoral and postdoctoral trainees. Currently, postdoctoral stipends at set based upon years of experience starting at 0 and ranging to 7 or more. The NIH notice number for 2021 was NOT-OD-21-049. Historical NIH NRSA salary data is available through the NIH website (\url{https://researchtraining.nih.gov/}).

\paragraph{Consumer Price Index (CPI)}
The Consumer Price Index (CPI) is a measure of the change of prices over time paid by consumers for a weighted set of goods and services. The CPI is released by the US Bureau of Labor Statistics (BLS) and is available through the BLS website (\url{https://www.bls.gov/cpi/}). 

\paragraph{Faculty hiring dataset}
The faculty hiring dataset was acquired from Clauset et al., \cite{clauset2015systematic}. The dataset consists of close to 19k faculty hires and their PhD training institution for Computer Science, History, and Business school faculty. We used the subset that completed their PhDs in the US. 

\paragraph{Fair Market Rents}
The Fair Market Rents dataset estimates the 40th percentile gross rents for standard quality units within a metropolitan or non-metropolitan statistical area. the data are provided by the US Office of Policy Development and Research (PD\&R). We used the 2-bedroom apartments data for this paper. 

\paragraph{NSF Survey of Earned Doctorates}
The NSF produces an annual Survey of Earned Doctorates which tracks information on educational history, demographics, and postgraduate plans after achieving a PhD. We used this survey to estimate the proportion of PhDs produced by each American university and city. Because the proportion of PhDs produced is used in comparison with the proportion of TT faculty dataset from Clauset et al., \cite{clauset2015systematic}, collected in 2013, we used the 2013 Survey of Earned Doctorates. 

\subsection{Postdoc salary processing}

From the full dataset of H1B applications we subsetted all postdoctoral positions (jobs titles containing the word "postdoctoral"). In total, the dataset has 34,804 salaries. From this dataset, we match employers in H1B applications to universities. We can match \statistic{28,217} applications to their corresponding universities. The remaining data were from industry, national laboratory, research foundation, and research hospital postdocs. We additionally removed \statistic{645} salaries where salaries are outside of the expected range (either less than \$30k or greater than \$100k) yielding a dataset of \statistic{27,572} salaries.

Real salaries were estimated using the same method as the BEA uses to estimate real income, without using a balancing factor. 

$$
Real\ Salary =  \frac{Salary}{RPP/100}
$$

\subsubsection{University inclusion criteria}
For university-level analyses (e.g. Fig \ref{fig:col}) we only included universities where we had at least 25 salaries for that university. 

\subsection{Adjustments for inflation}

Adjustments for inflation used the US city average CPI. In all cases throughout the manuscript where years are not listed for dollar values, 2021 dollars are used. 

\subsection{Data Availability}
All the data presented in this paper are publicly available online in the locations listed in Section \ref{sec:datasets}. Code required to reproduce these results are available at \url{https://github.com/timsainb/postdoc_salaries}. 

\end{document}